\documentclass[12pt,a4paper]{article}
\usepackage[latin9]{inputenc}
\usepackage{times}
\usepackage{a4wide}
\usepackage{amssymb}
\usepackage{amsmath}
\usepackage{bbm}
\usepackage{dsfont} 
\usepackage{ulem} 
\usepackage{pifont} 
\usepackage{ifpdf}
\ifpdf
\usepackage[pdftex]{graphicx}
\usepackage[pdftex,unicode,implicit]{hyperref}
\hypersetup{%
  pdftitle    = {The Tensor Hierarchies of Pure N=2,d=4,5,6 Supergravities},
  pdfkeywords = {N=2 Supersymmetry, supersmmetry algebra, gauge, symmetry},
  pdfauthor   = {M. Huebscher, T. Ortin and C.S. Shahbazi},
  pdfcreator  = {pdf\LaTeXe\ with package \flqq hyperref\frqq},
  pdfproducer = {pdf\LaTeXe\ with package \flqq hyperref\frqq},
  pdfpagemode = None,  
  pdffitwindow= true,  
  unicode     = true,
  plainpages  = false,
  colorlinks  = true,  
  citecolor   = blue,
  urlcolor    = red,
  linkcolor   = black
}
\newcommand{\hepth}[1]{({\tt \href{http://www.arXiv.org/abs/hep-th/#1}{hep-th/#1}})}

\newcommand{\arxiv}[1]{{\tt \href{http://www.arXiv.org/abs/#1}{arXiv:#1}}}
\else
  \usepackage[dvips]{graphicx}
  \usepackage[unicode,implicit]{hyperref}
  \newcommand{\hepth}[1]{{\tt hep-th/#1}}

  \newcommand{\arxiv}[1]{{\tt arXiv:#1}}
\fi
\makeatletter
\@addtoreset{equation}{section}
\makeatother

\begin{document}
\begin{flushright}
\small
IFT-UAM/CSIC-10-41\\
\texttt{arXiv:1006.4457}\\
October $18^{\rm th}$, $2010$
\normalsize
\end{flushright}
\begin{center}
\vspace{2cm}
{\LARGE {\bf The Tensor Hierarchies of Pure}}\\[.7cm]
{\LARGE {\bf  $N=2,d=4,5,6$ Supergravities}}
\vspace{1.8cm}

\begin{center}

 {\sl\large M.~H\"{u}bscher}
 \footnote{E-mail: {\tt huebscher@benoist-company.com}}$^{\dagger}$,
{\sl\large T.~Ort\'{\i}n}
\footnote{E-mail: {\tt Tomas.Ortin@cern.ch}}$^{\ddagger}$
 {\sl\large and  C.S.~Shahbazi}
 \footnote{E-mail: {\tt Carlos.Shabazi@uam.es}}$^{\ddagger}$

\vspace{.8cm}

$^{\dagger}${\it Benoist \& Company, Seefelder Str. 15, 82229 Seefeld,
Germany}

\vspace{.3cm}

$^{\ddagger}${\it Instituto de F\'{\i}sica Te\'orica UAM/CSIC\\
Facultad de Ciencias C-XVI,  C.U.~Cantoblanco,  E-28049-Madrid, Spain}\\

\end{center}

\vspace{1cm}


\textbf{Abstract}

\end{center}

\begin{quotation}\small
  We study the supersymmetric tensor hierarchy of pure (gauged) $N=2,d=4,5,6$
  supergravity and compare them with those of the pure, ungauged, theories
  (worked out in Ref.~\cite{Gomis:2007gb} for $d=5$) and the predictions of
  the Ka\v{c}-Moody approach made in Ref.~\cite{Kleinschmidt:2008jj}. We find
  complete agreement in the ungauged case but we also find that, after
  gauging, new St\"uckelberg symmetries reduce the number of independent
  \textit{physical} top-forms. The analysis has to be performed to all orders
  in fermion fields.

  We discuss the construction of the worldvolume effective actions for the
  $p$-branes which are charged with respect to the $(p+1)$-form potentials and
  the relations between the tensor hierarchies and $p$-branes upon dimensional
  reduction.
\end{quotation}

\newpage
\pagestyle{plain}


\newpage

\section*{Introduction}

The embedding-tensor formalism
\cite{Cordaro:1998tx,deWit:2005ub,deWit:2008ta,Bergshoeff:2009ph} has proven
to be a powerful method to determine, combining the requirements of duality
and gauge invariance, (a large part of) the $(p+1)$-form content
(\textit{tensor hierarchy}) of field theories and, in particular, of
supergravity theories. Thus, using the general results on 4-dimensional tensor
hierarchies of Ref.~\cite{Bergshoeff:2009ph}, the supersymmetric tensor
hierarchy of $N=1,d=4$ supergravity (and, implicitly, the most general form of
the $N=1,d=4$ theories) was found in Ref.~\cite{Hartong:2009az}. It turned out
to contain more forms than expected according to the arguments of
Ref.~\cite{Bergshoeff:2009ph}, based on (bosonic) gauge symmetry, but it is
still possible to make sense of all of them since the new 3-forms could be
associated to new deformations of the theory and the new 4-forms to new
constraints involving the embedding tensor and other deformation
parameters. The 3-forms transform into the gravitino under supersymmetry and
they were shown to be associated to supersymmetric domain-wall solutions in
Ref.~\cite{Huebscher:2009bp}.

At the same time, it has been conjectured that the $(p+1)$-form spectra
different supergravity theories are determined by very extended symmetry groups
related to their duality groups (see, e.g.~Ref.~\cite{Kleinschmidt:2008jj} and
references therein). It is not guaranteed that these two approaches (the
embedding tensor approach and the \textit{Ka\v{c}-Moody} (KM) approach) will
give the same $(p+1)$-form spectra and, in order to find all the $(p+1)$-forms
of a given supergravity (and all the $p$-branes associated to them) it is
important to compare the predictions of both methods. In this paper we will do
this for the simplest $N=2,d=4,5,6$ theories.

Our first goal will be to extend the results obtained for the $N=1,d=4$
theories in \cite{Hartong:2009az} to the $N=2,d=4,5,6$ theories focusing, for
simplicity, on the pure supergravity theories (i.e.~in absence of matter
couplings). We will, therefore, find the supersymmetric tensor hierarchies of
these theories using the embedding-tensor formalism and using the results in
Refs.~\cite{Gomis:2007gb,Kleinschmidt:2008jj} and \cite{Hartong:2009vc}. We
will then compare the spectra of $(p+1)$-forms obtained with the predictions
of the KM approach made in Ref.~\cite{Kleinschmidt:2008jj}.

The absence of matter couplings limits the range of $(p+1)$-forms that we can
find. However, those associated to the R-symmetry of the theory (not
considered in Ref.~\cite{de Vroome:2007zd}), which transform into gravitini
(the only fermions available in the pure $N=2$ theories), and which,
therefore, may be associated to dynamical $p$-branes, should not be missed in
our analysis. We will check this correspondence by constructing explicitly
with the $(p+1)$-forms of the tensor hierarchies candidates to the bosonic
part of $\kappa$-symmetric worldvolume actions\footnote{Some previous partial
  results were also given in Ref.~\cite{Bergshoeff:2007ij}.}.

This paper is organized as follows: in Section~\ref{eq:N2sugra} we give our
conventions for the pure, ungauged, $N=2,d=4$ supergravity theory and in
Section~\ref{sec-hierarchy} we construct its tensor hierarchy as a particular
case of the generic (bosonic) 4-dimensional hierarchy of
Ref.~\cite{Bergshoeff:2009ph}, on the basis of the R-symmetry group of the
theory. In Section~\ref{sec-N2d4sugramagnetic} we consider the gauging of the
R-symmetry group, constructing the supersymmetry transformation rules for all
the $(p+1)$-form fields in the tensor hierarchy to lowest order in fermions
and checking their closure up to duality relations. In
Section~\ref{sec-5dtheory} we briefly review the analogous results for the
minimal $d=5$ supergravity found in
Ref.~\cite{Gomis:2007gb,Kleinschmidt:2008jj}, from the point of view of the
5-dimensional tensor hierarchy of Ref.~\cite{Hartong:2009vc} and in
Section~\ref{sec-6dtheory} we cover the 6-dimensional case, which is much
simpler due to the absence of 1-forms. In
Section~\ref{sec-puren2objectsandactions} we discuss the construction of the
effective actions for the $p$-branes of these theories and in
Section~\ref{sec-conclusions} we present our conclusions.


\section{Pure, ungauged,  $N=2,d=4$ supergravity}
\label{eq:N2sugra}

In order to pave the way for further generalizations, we are going to describe
pure $N=2,d=4$ supergravity as a particular case of the general matter-coupled
$N=2,d=4$ supergravity\footnote{See
  Refs.~\cite{Andrianopoli:1996cm,kn:toinereview} and the original references
  \cite{deWit:1984pk,deWit:1984px}. Our conventions are given in
  Refs.~\cite{Bellorin:2005zc,Meessen:2006tu,Huebscher:2006mr} and follow
  closely those of Ref.~\cite{Andrianopoli:1996cm}. In particular, our
  $\sigma$ matrices satisfy
\begin{equation}
\label{eq:productofPaulimatrices}
(\sigma^{x}\sigma^{y})^{I}{}_{J} \; =\; \delta^{xy}\ \delta^{I}{}_{J} 
+ i\varepsilon^{xyz}\ (\sigma^{z})^{I}{}_{J}\, ,
\end{equation}
and we  \textit{define}
\begin{equation}
(\sigma^{x})_{I}{}^{J} \equiv [(\sigma^{z})^{I}{}_{J} ]^{*}\, , 
\end{equation}
and one finds that
\begin{equation}
(\sigma^{x})_{I}{}^{J}
=\varepsilon_{IK}(\sigma^{z})^{K}{}_{L}\varepsilon^{LJ}\, ,
\hspace{.5cm}
\sigma^{x\, [I}{}_{J}\varepsilon^{K]J}
=\sigma^{x}{}_{[I}{}^{K}\varepsilon_{J]K}=0\, 
\hspace{.5cm}
\sigma^{x}{}_{I}{}^{J} = \sigma^{x\, J}{}_{I}\, .
\end{equation}
}. To make contact with the
conventions used in the embedding-tensor formalism we will write symplectic
products with symplectic indices $M,N$.

The supergravity multiplet of the $N=2,d=4$ theory consists of the graviton
$e^{a}{}_{\mu}$, a pair of gravitini $\psi_{I\, \mu}\, ,\,\,\, (I=1,2)$ which
we describe as Weyl spinors, and one graviphoton which we denote by
$A^{\Lambda}{}_{\mu}$ even though the index $\Lambda$ only takes one value to
distinguish this fundamental (electric) field from its (magnetic) dual
$A_{\Lambda\, \mu}$. The bosonic action is

\begin{equation}
\label{eq:pureungaugedN2D4bosonicaction}
  S  =  -{\displaystyle\int} 
  \left[\star R 
    +4\Im{\rm m}\mathcal{N}_{\Lambda\Sigma} 
    F^{\Lambda}\wedge \star F^{\Sigma}
    +4\Re{\rm e}\mathcal{N}_{\Lambda\Sigma} 
    F^{\Lambda}\wedge F^{\Sigma}
  \right]\, ,
\end{equation}
 
\noindent
where $\mathcal{N}_{\Lambda\Sigma}$ is the period ``matrix'' with only one
component with negative imaginary part. In this case the choice of period
``matrix'' (or, equivalently, of constant ``canonical symplectic section''
$\mathcal{V}^{M}$) is arbitrary, as far as the constraints are satisfied.

We define $G_{\Lambda}$ by 

\begin{equation}
\label{eq:GL}
G_{\Lambda}{}^{+} = \mathcal{N}^{*}_{\Lambda\Sigma}F^{\Lambda\, +}\, ,  
\end{equation}

\noindent
and define the 2-dimensional symplectic vector

\begin{equation}
\label{eq:GMdef}
(G^{M}) \equiv  
\left(
  \begin{array}{c}
    F^{\Lambda} \\
    G_{\Lambda} \\ 
  \end{array}
\right)\, .
\end{equation}

The supersymmetry transformations of the supergravity fields to all orders in
fermions are

\begin{eqnarray}
\label{eq:susytranse}
\delta_{\epsilon} e^{a}{}_{\mu} & = & 
-{\textstyle\frac{i}{4}} \bar{\psi}_{I\, \mu}\gamma^{a}\epsilon^{I}
+\mathrm{c.c.}\, ,\\
& & \nonumber \\ 
\label{eq:gravisusyrule}
\delta_{\epsilon}\psi_{I\, \mu} & = & 
\tilde{\nabla}_{\mu}\epsilon_{I} 
+\varepsilon_{IJ}\tilde{T}^{+}{}_{\mu\nu}\gamma^{\nu}\epsilon^{J}
\, ,\\
& & \nonumber \\ 
\label{eq:susytransA}
\delta_{\epsilon} A^{\Lambda}{}_{\mu} & = & 
{\textstyle\frac{1}{4}}
\mathcal{L}^{\Lambda\, *}
\varepsilon^{IJ}\bar{\psi}_{I\, \mu}\epsilon_{J}
+\mathrm{c.c.}\, ,
\end{eqnarray}

\noindent
where $\tilde{\nabla}_{\mu}$ is the Lorentz covariant derivative that uses the
torsionful spin connection $\tilde{\omega}_{\mu}{}^{ab}$:

\begin{equation}
\label{eq:torsionfuld4}
  \begin{array}{rcl}
\tilde{\omega}_{abc} & = & 
-\tilde{\Omega}_{abc} +\tilde{\Omega}_{bca} -\tilde{\Omega}_{cab}\, ,
\\
& & \\
\tilde{\Omega}_{abc} & = & \Omega_{abc} +\tfrac{1}{2}T_{abc}\, ,
\\
& & \\
\Omega_{abc} & = & e_{a}{}^{\mu}e_{b}{}^{\nu}\partial_{[\mu|}e_{c|\nu]}\, ,
\\
& & \\
T_{\mu\nu}{}^{a} & = & -\tfrac{i}{2}\bar{\psi}_{[\mu|\,
  I}\gamma^{a}\psi_{|\nu]}{}^{I}\, ,
  \end{array}
\end{equation}

\noindent
$\mathcal{L}^{\Lambda}$ is the upper component of the
canonically-normalized symplectic section 

\begin{equation}
\label{eq:conicalnorm}
(\mathcal{V}^{M}) = 
\left(
  \begin{array}{c}
\mathcal{L}^{\Lambda} \\
\mathcal{M}_{\Lambda} \\
\end{array}
\right)\, ,
\hspace{1cm}
\langle \mathcal{V}\mid\mathcal{V}^{*}\rangle 
\equiv  
\mathcal{V}^{M\, *}\mathcal{V}_{M}
=
\mathcal{L}^{*\, \Lambda}\mathcal{M}_{\Lambda} 
-\mathcal{L}^{\Lambda}\mathcal{M}^{*}_{\Lambda}
= -i\, ,  
\end{equation}

\noindent
and where $\tilde{T}$ is the supercovariant graviphoton field strength which
can be written in the form

\begin{equation}
  \tilde{T}^{+}  =  \langle\,\mathcal{V} \mid \tilde{F}^{+}\,\rangle = 
\tilde{G}^{M\, +}\mathcal{V}_{M}\, ,
\end{equation}

\noindent
$\tilde{F}^{\Lambda}$ being given by

\begin{equation}
\tilde{F}^{\Lambda}{}_{\mu\nu} = F^{\Lambda}{}_{\mu\nu} 
+\tfrac{1}{4}
\left[\mathcal{L}^{\Lambda}\varepsilon_{IJ}\bar{\psi}^{I}{}_{\mu}\psi^{J}{}_{\nu}
+\mathrm{c.c.}\right]\, .  
\end{equation}

A convenient choice of symplectic section and period matrix, satisfying the
relations

\begin{equation}
\label{eq:LL}
\mathcal{M}_{\Lambda}=\mathcal{N}_{\Lambda\Sigma}\mathcal{L}^{\Sigma}\, ,
\hspace{1cm}
\mathcal{L}^{*\Lambda}  \mathcal{L}^{\Sigma} = -\tfrac{1}{2}\Im{\rm m}\,
\mathcal{N}^{\Lambda\Sigma}\, ,
\end{equation}

\noindent
(where the upper indices in the period matrix indicate that we are dealing
with the inverse) is\footnote{It can be shown that the most general
  $\mathcal{V}^{M}$ satisfying the normalization constraint can always be
  brought to this one by a symplectic transformation.}

\begin{equation}
\label{eq:convenientVM}
\mathcal{V} = 
(\mathcal{V}^{M}) = 
\left(
  \begin{array}{c}
    i \\ {\textstyle\frac{1}{2}} \\
  \end{array}
\right)\, ,  
\hspace{1cm}
\mathcal{N}_{\Lambda\Sigma}=-\frac{i}{2}\, ,
\end{equation}

\noindent
but we will leave them undetermined in what follows and we will only use
their constancy and their general properties.


\section{The tensor hierarchy of pure $N=2, d=4$ supergravity}
\label{sec-hierarchy}

In this section we are going to determine the tensor hierarchy of pure $N=2,
d=4$ supergravity by adapting the generic result of
Ref.~\cite{Bergshoeff:2009ph} to the actual global symmetries and field
content of the theory at hands.


\subsection{Global symmetries of the ungauged theory}
\label{sec-symm}

The bosonic global symmetry of this theory is the $Sp(2)$ group of
electric-magnetic duality rotations of the vector fields that preserve the
equations of motion but not the action:

\begin{equation}
\delta_{\alpha^{\rm a}} G^{M} = \alpha^{\rm a} T_{{\rm a}\, N}{}^{M}G^{N}\, , 
\end{equation}

\noindent
where the generators $T_{{\rm a}\, N}{}^{M}$ satisfy the algebra

\begin{equation}
[T_{\rm a},T_{\rm b}]=-f_{\rm ab}{}^{\rm c}T_{\rm c}\, ,
\hspace{1cm}
f_{12}{}^{3}=-f_{23}{}^{1}=-f_{31}{}^{2}=-1\, .  
\end{equation}

\noindent
This symmetry cannot be gauged using the electric and magnetic vector fields
$A^{M}$ because they are charged under it\footnote{One can only expect to be
  able to gauge one of the Abelian subgroups (there is no non-Abelian
  2-dimensional subgroup) with just two gauge fields. It is easy to see that
  this is impossible.}. One can also see that the quadratic constraint of the
embedding-tensor formalism cannot be satisfied for $\vartheta_{M}{}^{\rm
  a}\neq 0$.

The other global symmetry of the theory is the R-symmetry group
$U(2)=U(1)\times SU(2)$ which only acts on the spinors of the theory according
to

\begin{equation}
\delta \psi_{\mu\, I} = \alpha^{\mathbf{x}}T_{\mathbf{x}\, I}{}^{J}\psi_{\mu\,
  J}\, ,   
\end{equation}

\noindent
where 

\begin{equation}
T_{\mathbf{x}\, I}{}^{J} = -\tfrac{i}{2}\sigma^{\mathbf{x}}{}_{I}{}^{J}\, ,  
\hspace{1cm}
\mathbf{x} =0,x\, ,
\hspace{1cm}
x=1,2,3\, ,
\end{equation}

\noindent
where $\sigma^{x}{}_{I}{}^{J}$s are the Pauli matrices and
$\sigma^{0}{}_{I}{}^{J}\equiv \delta_{I}{}^{J}$. The non-vanishing commutators
between these generators are\footnote{We will not distinguish between upper
  and lower $SU(2)$ indices $x,y,z$.}

\begin{equation}
[T_{x},T_{y}]_{I}{}^{J} = -\varepsilon_{xyz}T_{z\, I}{}^{J}\, ,    
\end{equation}

\noindent
so, in the conventions of Ref.~\cite{Bergshoeff:2009ph}, the structure
constants are $f_{xy}{}^{z}=+\varepsilon_{xyz}$.  The $\mathbf{x}=0$
transformations are just multiplication by a (``K\"ahler'') $U(1)_{\rm
  Kahler}$ phase to be distinguished from the $U(1)$ subgroups of $SU(2)$.

The standard quadratic constraint that expresses the invariance of the
embedding tensor takes for this symmetry the form

\begin{equation}
\vartheta_{M}{}^{x} \vartheta_{N}{}^{y}\varepsilon_{xyz}=0\, ,
\end{equation}

\noindent
i.e.~the two 3-component vectors $\vartheta^{\Lambda\, x}$ and
$\vartheta_{\Lambda}{}^{x}$ are parallel and we can write

\begin{equation}
\vartheta_{M}{}^{x} = \alpha_{M}\vartheta^{x}\, ,  
\end{equation}

\noindent
where $\alpha_{M}$ is an arbitrary 2-component symplectic vector.  

The quadratic constraint $\vartheta^{M\, [x}\vartheta_{M}{}^{y]}=0$ is
automatically satisfied. The quadratic constraint $\vartheta^{M\,
  [x}\vartheta_{M}{}^{0]}=0$ is satisfied if
$\vartheta_{M}{}^{0}=\vartheta^{0} \alpha_{M}$

$\vartheta^{x}$ selects the $U(1)\subset SU(2)$ which is to be gauged combined
with $U(1)_{\rm Kahler}$. The generator of the Abelian symmetry which is
gauged is, therefore, $\vartheta^{\mathbf{x}}T_{\mathbf{x}}$. The vector
$\alpha_{M}$ selects a combination of the two vector fields $\alpha_{M}A^{M}$
that will act as a gauge vector.

It is well known that $U(1)_{\rm Kahler}$ cannot be gauged in $N=2,d=4$
supergravity and, therefore, we will set $\vartheta^{0}= 0$ from the onset.
Furthermore, since the electric-magnetic duality group $Sp(2)$ cannot be
gauged, we are also going to set $T_{\rm a}=0$ from the
beginning\footnote{There is no way to define consistent supersymmetry rules
  for the corresponding 2-forms $B_{\rm a}$ anyway.}.


\subsection{The tensor hierarchy of pure $N=2, d=4$ supergravity}
\label{sec-bosonicTH}

In principle, we can naively substitute the indices of the symmetries, 1-forms
and embedding tensor involved in our problem into the generic formulae of
Ref.~\cite{Bergshoeff:2009ph}. We only have to consider the R-symmetry with
$f_{xy}{}^{z}=+\varepsilon_{xyz}$ and $\vartheta_{M}{}^{x} =
\alpha_{M}\vartheta^{x}$, which leads to $X_{MNP}=0$.



We should also take into account that both the 4-form $D_{x}{}^{NM}$ and the
associated 3-form gauge parameter $\Lambda_{x}{}^{NM}$ can be taken to be
antisymmetric in the upper indices. Then, we can define

\begin{eqnarray}
D^{(1)}{}_{x} 
& \equiv & 
\tfrac{1}{2}\Omega_{PQ} D_{x}{}^{PQ}\, ,
\\
& & \nonumber \\
D^{(2)}{}_{x} 
& \equiv & 
-\tfrac{1}{8}\varepsilon_{xyz} D_{yz}\, ,
\\
& & \nonumber \\
\Lambda^{(1)}{}_{x} 
& \equiv & 
\tfrac{1}{2}\Omega_{PQ} \Lambda_{x}{}^{PQ}\, ,
\\
& & \nonumber \\
\Lambda^{(2)}{}_{x} 
& \equiv & 
-\tfrac{1}{8}\varepsilon_{xyz} \Lambda_{yz}\, ,
\end{eqnarray}

\noindent
Furthermore, the triplets $D^{(1)}_{x},D^{(2)}_{x}$ only appear everywhere
through their sum, so their St\"uckelberg shifts with 4-form parameters
$\tilde{\Lambda}_{x}$ cancel each other. This indicates the existence of just
one independent triplet of 4-forms and we define\footnote{Alternatively, we
  may use the St\"uckelberg shift to eliminate one of these triplets.}

\begin{equation}
D_{x}\equiv   D^{(1)}{}_{x}+D^{(2)}{}_{x}\, ,
\hspace{1cm}
\Lambda_{x}\equiv \Lambda^{(1)}{}_{x}+\Lambda^{(2)}{}_{x}\, .
\end{equation}

\noindent
Finally, we find that $D_{x}$ can have additional gauge transformations that
leave invariant the 4-form field strength $G_{x}{}^{M}$:

\begin{equation}
\label{eq:extrastuckelberg4form}
\delta D_{x} =\vartheta^{x}\tilde{\Lambda}\, ,   
\end{equation}

\noindent
where $\tilde{\Lambda}$ is a 4-form. This St\"uckelberg shift cannot be used
to eliminate the complete triplet of 4-forms $D_{x}$, but only a combination
of them.

We find that, out of all the possible fields present in the generic tensor
hierarchy, $A^{M},B_{x},C_{x}{}^{M}$ and $D_{x}$ are interconnected by gauge
transformations, while $B_{0},C_{0}{}^{M},D_{0x},D_{0}{}^{NM}$ and $D^{NPQ}$
are decoupled from them. We can advance that we have been able to construct
consistent supersymmetry transformation rules for all the fields in the first
group plus $B_{0}$ (as in Ref.~\cite{Bergshoeff:2007ij}) but not for the rest
of the fields in the second group. Thus we will ignore them from now on
although in presence of matter they might be coupled consistently with the
rest of the theory.

Taking all this information into account, we find the gauge transformations

\begin{eqnarray}
\label{eq:dhAM}
\delta_{h} A^{M} 
& =  & 
-d\Lambda^{M} +\tfrac{1}{2}\alpha^{M}\vartheta^{x}\Lambda_{x}\, ,
\\
& & \nonumber \\  
\label{eq:deltahBx}
\delta_{h} B_{x} 
& = &  
\mathfrak{D}\Lambda_{x}  
-\varepsilon_{xyz}\vartheta^{y}\Lambda_{z}{}^{M}\alpha_{M}\, ,
\\
& & \nonumber \\
\delta_{h} C_{x}{}^{M} 
& = & 
\mathfrak{D}\Lambda_{x}{}^{M}
-F^{M}\wedge\Lambda_{x}
-\delta_{h} A^{M}\wedge B_{x}
+\Lambda^{M}H_{x}
-2\alpha^{M} \vartheta^{y}\varepsilon_{xyz}\Lambda_{z}\, ,
\\
& & \nonumber \\
\label{eq:deltahDx1+Dx2}
\delta_{h} D_{x} 
& = & 
\mathfrak{D}\Lambda_{x}
+\tfrac{1}{2}[F_{M}
+{\textstyle\frac{1}{4}}\alpha_{M}\vartheta^{y}B_{y}]\wedge
\Lambda_{x}{}^{M}
 +\tfrac{1}{2}\delta_{h} A_{M}\wedge C_{x}{}^{M}
\nonumber \\
& & \nonumber \\
& & 
-\tfrac{1}{2}\Lambda_{M}G_{x}{}^{M}
-\tfrac{1}{8}\varepsilon_{xyz}\mathfrak{D}\Lambda_{y}\wedge B_{z} 
+\tfrac{1}{4}\varepsilon_{xyz}\Lambda_{y}\wedge H_{z}
+\vartheta^{x}\tilde{\Lambda}\, .  
\end{eqnarray}

\noindent
where the $SU(2)$-covariant derivatives are given, e.g., by

\begin{equation}
\mathfrak{D}\Lambda_{x}  =  d\Lambda_{x}  
+\varepsilon_{xyz}\vartheta^{y}\alpha_{M}A^{M}\wedge \Lambda_{z}\, .
\end{equation}

\noindent 
We also find the gauge-covariant field strengths

\begin{eqnarray}
\label{eq:FM}
F^{M} 
& = & 
dA^{M} -\tfrac{1}{2}\alpha^{M}\vartheta^{x}B_{x}\, ,
\\
& & \nonumber \\
\label{eq:Hx}
H_{x} 
& = & 
\mathfrak{D}B_{x}  +\alpha_{M}\varepsilon_{xyz}\vartheta^{y}C_{z}{}^{M}\, , 
\\
& & \nonumber \\
\label{eq:GxM}
G_{x}{}^{M} 
& = & 
\mathfrak{D}C_{x}{}^{M}
+[F^{M}+{\textstyle\frac{1}{4}}\alpha^{M}\vartheta^{y}B_{y}]\wedge B_{x}
+2\alpha^{M}\varepsilon_{xyz}\vartheta^{y}D_{z}\, , 
\end{eqnarray}

\noindent
and the hierarchical Bianchi identities

\begin{eqnarray}
dF^{M} 
& = & 
-\tfrac{1}{2}\alpha^{M}\vartheta^{x}H_{x}\, ,
\\
& & \nonumber \\
\mathfrak{D}H_{x} 
& = & 
\alpha_{M}\varepsilon_{xyz} \vartheta^{y}G_{z}{}^{M}  \, .
\end{eqnarray}

For the decoupled 2-form $B_{0}$ we find trivial results

\begin{equation}
\delta_{h} B_{0} 
 =   
d\Lambda_{0}\, ,
\hspace{.5cm}
H_{0} 
 =  
dB_{0}\, , 
\hspace{.5cm}
dH_{0}  = 0\, .
\end{equation}

This is the tensor hierarchy that naively follows from the general one.
However, we observe that by setting $\vartheta^{0}=0$ we have removed $B_{0}$
from $F^{M}$, for instance. This decoupling allows for more general gauge
transformations for $B_{0}$ that cannot be determined by the embedding-tensor
method and will be determined by supersymmetry.


\section{General gauging of pure $N=2, d=4$ supergravity}
\label{sec-N2d4sugramagnetic}

We want to gauge the global symmetries of the theory discussed in the previous
section using as gauge fields the electric and magnetic graviphoton, using the
embedding-tensor formalism\footnote{The electrically-gauged theory was
  constructed in \cite{Freedman:1976aw,Fradkin:1976xz}.}. We have seen that
gauge invariance requires higher-rank fields but their presence must be
compatible with supersymmetry.
 
We are going to proceed as in Ref.~\cite{Hartong:2009az}: first of all, we
make the following electric-magnetic invariant Ansatz for the gravitini
supersymmetry transformations\footnote{Symplectic indices are raise and
  lowered with the symplectic metric according to
  $\mathcal{V}_{M}=\mathcal{V}^{N}\Omega_{MN}$ and
  $\mathcal{V}^{M}=\Omega^{NM}\mathcal{V}_{N}$, where
  $\Omega_{MN}=\Omega^{MN}$ and $\Omega^{MP}\Omega_{NP}=\delta^{M}{}_{N}$.}
to lowest order in fermions:

\begin{equation}
\delta_{\epsilon} \psi_{\mu\, I} =
\tilde{\mathfrak{D}}_{\mu} \epsilon_{I}
+\epsilon_{IJ}\mathcal{V}_{M}\tilde{F}^{M\, +}{}_{\mu\nu}\gamma^{\nu}\epsilon^{J}
-\tfrac{1}{4}\mathcal{V}^{M}\alpha_{M}\vartheta^{x} 
\varepsilon_{IK}\sigma^{x\, K}{}_{J}\gamma_{\mu}\epsilon^{J}\, ,
\end{equation}

\noindent
where 

\begin{equation}
  \tilde{\mathfrak{D}}_{\mu} \epsilon_{I}
  =
  \tilde{\nabla}_{\mu} \epsilon_{I}
  +\tfrac{i}{2}A^{M}{}_{\mu}\alpha_{M}\vartheta^{x} 
  \sigma^{x}{}_{I}{}^{J}\epsilon_{J}\, ,  
\end{equation}

\noindent
is the Lorentz and gauge-covariant derivative of the supersymmetry parameter
$\epsilon_{I}$ that uses the torsionful spin connection in
Eq.~(\ref{eq:torsionfuld4}) and where $\tilde{F}^{M}$ is the
supercovariantization of the vector field strength of the tensor hierarchy of
the tensor hierarchy defined in Eq.~(\ref{eq:FM})

\begin{equation}
\tilde{F}^{M}{}_{\mu} \equiv   
F^{M}{}_{\mu\nu} 
+\tfrac{1}{4}
\left[\mathcal{V}^{M}\varepsilon_{IJ}\bar{\psi}^{I}{}_{\mu}\psi^{J}{}_{\nu}
+\mathrm{c.c.}\right]\, .
\end{equation}

\noindent
It should be distinguished from $\tilde{G}^{M}$, defined in
Eq.~(\ref{eq:GMdef}), because the latter only depends on the electric vector
field $A^{\Lambda}$.

This supersymmetry transformation reduces to the standard one for purely
electric gaugings upon use of the duality relation
$\tilde{F}_{\Lambda}=\tilde{G}_{\Lambda}$ which we can also write

\begin{equation}
\label{eq:FM=GM}
\tilde{F}^{M}=\tilde{G}^{M}\, .
\end{equation}

The supersymmetry variations of the bosonic fields (Vierbein and electric
graviphoton) are not modified, but we have to add a supersymmetry
transformation for the magnetic vector field $A_{\Lambda}$ compatible with
symplectic symmetry. A symplectic-covariant Ansatz that gives correctly the
supersymmetry transformation rule for the electric graviphoton
Eq.~(\ref{eq:susytransA}) when $\mathcal{V}^{M}$ is given by
Eq.~(\ref{eq:convenientVM}) and also coincides with the uncoupled case of
Ref.~\cite{Bergshoeff:2007ij}) is

\begin{equation}
\label{eq:susytransAM}
\delta_{\epsilon} A^{M}{}_{\mu} =
-{\textstyle\frac{1}{4}}\mathcal{V}^{M\, *}\epsilon^{IJ}
\bar{\epsilon}_{I} \psi_{\mu\, J}+\mathrm{c.c.}
\end{equation}

The local supersymmetry algebra acting on $A^{M}{}_{\mu}$ closes into

\begin{equation}
[\delta_{\eta},\delta_{\epsilon}]
A^{M}{}_{\mu}
=  
[\delta_{\rm g.c.t.}(\xi)  +\delta_{h}(\Lambda^{M},\Lambda_{x}) +\delta_{\rm susy}(\kappa)]
A^{M}{}_{\mu}\, ,
\end{equation}

\noindent
where $\delta_{\rm g.c.t.}(\xi)$ is a general coordinate transformation with
parameter $\xi^{\mu}$, $\delta_{h}(\Lambda^{M},\Lambda_{x})$ is the gauge
transformation predicted by the tensor hierarchy given in Eq.~(\ref{eq:dhAM})
with parameters $\Lambda^{M},\Lambda_{x}$ and $\delta_{\rm susy}(\kappa)$ is
the supersymmetry transformation of Eq.~(\ref{eq:susytransAM}) with parameter
$\kappa$ if a term of the form

\begin{equation}
F^{M}{}_{\nu\mu} +2\Im{\rm m}(\mathcal{V}^{*M}\mathcal{V}_{N}\star
F^{N+}{}_{\nu\mu}) 
\end{equation}

\noindent
vanishes. This term, upon use of Eq.~(\ref{eq:bigidentity}), takes the form
of the duality relation Eq.~(\ref{eq:FM=GM}).

The parameters $\xi,\Lambda^{M},\Lambda_{x},\kappa$ are given by the
spinor bilinears

\begin{eqnarray}
\xi^{\mathbf{x}}{}_{\mu} 
& \equiv & 
{\textstyle\frac{i}{4}}\sigma^{\mathbf{x}}{}_{J}{}^{I} 
(\bar{\epsilon}_{I}\gamma_{\mu}\eta^{J}
-\bar{\eta}_{I}\gamma_{\mu}\epsilon^{J})
\, \in\, \mathbb{R}\, ,
\hspace{1cm}
\xi^{0}{}_{\mu} \equiv \xi_{\mu}\, ,
\\
& & \nonumber \\
X 
& \equiv & 
{\textstyle\frac{1}{2}}\epsilon_{IJ}\bar{\epsilon}^{I}\eta^{J}\, ,
\\
& & \nonumber \\
\label{eq:lambdaM}
\Lambda^{M}
& = & 
\Re{\rm e}\, (\mathcal{V}^{M}X) +\xi^{\mu}A^{M}{}_{\mu}\, ,
\\
& & \nonumber \\
\label{eq:lambdax}
\Lambda_{x\, \mu}  
& = & 
-\tfrac{1}{2}\xi^{x}{}_{\mu}-\xi^{\nu}B_{x\, \nu\mu}\, .
\\
& & \nonumber \\
\label{eq:kappad4}
\kappa^{I}
& = & 
-\xi^{\mu}\psi^{I}{}_{\mu}\, .
\end{eqnarray}

The use of the vector field strengths $F^{M}$ containing a triplet of 2-forms
$B_{x\, \mu\nu}$ predicted by the tensor hierarchy in the gravitini
supersymmetry transformations is clearly justified due to the presence of
St\"uckelberg shifts of the vector fields.  Now, the consistency of this
construction requires the existence of a triplet of 2-forms $B_{x\, \mu\nu}$
with consistent supersymmetry transformations and with the gauge
transformations predicted by the tensor hierarchy.

Observe that the more general Ansatz

\begin{equation}
\label{eq:1formanstaz}
\alpha \mathcal{V}^{M} \sigma^{\mathbf{x}}{}_{IJ}
\bar{\epsilon}^{I}\psi_{\mu}{}^{J}
+\mathrm{c.c.}\, ,
\end{equation}

\noindent
which would include for a possible triplet of 1-forms not predicted by the
tensor hierarchy does not work for $\mathbf{x}\neq 0$.


\subsection{2-forms}

According to the tensor hierarchy predictions, we expect a set of 2-forms
$B_{\mathbf{x}\, \mu\nu}$, 3 of which are required by the consistency of the
supersymmetry transformations of $A^{M}$. Actually, we can only make Ansatze
for the supersymmetry transformation rules of as many 2-forms:

\begin{equation}
\label{eq:2formanstaz}
\delta_{\epsilon}B_{\mathbf{x}\, \mu\nu}
=\beta i \sigma^{\mathbf{x}}{}_{I}{}^{J}  
\bar{\epsilon}^{I}\gamma_{[\mu}\psi_{\nu] J}
+\mathrm{c.c.}\, ,
\end{equation}

\noindent
where the constant $\beta$ has to be a real for the local supersymmetry
algebra to close.  

For $\mathbf{x}=x$ we find that the transformation

\begin{equation}
\label{eq:susytransBx}
\delta_{\epsilon}B_{x\, \mu\nu}
=\tfrac{i}{4} \sigma^{x}{}_{I}{}^{J}  
\bar{\epsilon}^{I}\gamma_{[\mu}\psi_{\nu] J}
+\mathrm{c.c.}\, ,
\end{equation}

\noindent
leads to  closure of the local supersymmetry algebra 

\begin{equation}
\label{eq:commutatorBx}
\left[\delta_{\eta},\delta_{\epsilon}\right] B_{x\, \mu\nu} 
= 
[\delta_{\rm g.c.t.}(\xi) +\delta_{h}(\Lambda_{x}, \Lambda_{x}{}^{M})
+\delta_{\rm susy}(\kappa)] 
B_{x\, \mu\nu} \, ,
\end{equation}

\noindent
if we impose the duality (or on-shell) condition

\begin{equation}
\tilde{H}_{x}=0\, ,
\end{equation}

\noindent
where $\tilde{H}_{x}$ is the 3-form supercovariant field strength

\begin{equation}
\tilde{H}_{x\, \mu\nu\rho}
=
H_{x\, \mu\nu\rho}
-\tfrac{3i}{4}\sigma^{x}_{I}{}^{J}\bar{\psi}^{I}{}_{[\mu}\gamma_{\nu| }\psi_{J\,
  |\rho]} \, .
\end{equation}

\noindent
The parameters $\xi, \Lambda_{x}, \kappa$ are the spinor bilinears defined
before (as they must) and $\Lambda_{x}{}^{M}$ is given by

\begin{eqnarray}
\Phi^{x}{}_{\mu\nu} 
& \equiv & 
\sigma^{x\, K}{}_{J}\varepsilon_{IK}
\bar{\epsilon}^{I}\gamma_{\mu\nu}\eta^{J}\, ,  
\\
& & \nonumber \\
\Lambda_{x}{}^{M}{}_{\mu\nu}
& \equiv &
+\tfrac{1}{4}\Re{\rm e}\, 
(\mathcal{V}^{M} \Phi^{x}{}_{\mu\nu}) 
-a^{M}B_{x\, \mu\nu} -\xi^{\rho}C_{x}{}^{M}{}_{\rho\mu\nu}\, .
\end{eqnarray}

Observe that the three 2-forms $\Phi^{x}{}_{\mu\nu}$ are anti-selfdual.

Thus, we find complete agreement with the tensor hierarchy prediction and
consistency with the supersymmetry transformations proposed for
$A^{M}$. Furthermore, the on-shell condition $\tilde{H}_{x}=0$ can be
rewritten in the form

\begin{equation}
H_{x}
= \star j_{N\, x}\, ,
\end{equation}

\noindent
where

\begin{equation}
j_{N\, x}{}^{\mu} \equiv \tfrac{i}{8}\epsilon^{\mu\alpha\beta\gamma}   
\sigma^{x}_{I}{}^{J}\bar{\psi}^{I}{}_{\alpha}\gamma_{\beta}\psi_{J\,
  \gamma}\, ,
\end{equation}
 
\noindent
is the triplet of Noether currents associated to the global $SU(2)$
invariance, in agreement with the general arguments of
Ref.~\cite{Bergshoeff:2009ph}. At lowest order in fermion fields these Noether
currents vanish and instead of a duality condition the on-shell condition
$H_{x}$ resembles a gauge-triviality condition for the 2-forms.

For $\mathbf{x}=0$ we get, to lowest order in fermions

\begin{equation}
\left[\delta_{\eta},\delta_{\epsilon}\right] B_{0\, \mu\nu} 
=
2\partial_{[\mu}(-2\beta  \xi_{\nu]}) 
+8\beta \Im{\rm m}\, [X\mathcal{V}_{M} F^{M+}{}_{\mu\nu}]
+\beta\alpha_{M}\vartheta^{x} \Im{\rm m}[\mathcal{V}^{M} \Phi^{x}{}_{\mu\nu}] 
\, .
\end{equation}

\noindent
Now, using the identities\footnote{All these identities can be derived from
  the constraint $\mathcal{V}^{*\, M}\mathcal{V}_{M}=-i$ and
  Eq.~(\ref{eq:RR}).}

\begin{eqnarray}
\label{eq:ImRerelation}
\Im{\rm m}\, (\mathcal{V}^{M}X)  & = &  
-2 \Re{\rm e}\, (\mathcal{V}^{*\, M}\mathcal{V}_{N}) 
\Re{\rm e}\, (\mathcal{V}^{N}X)\, ,
\\
& & \nonumber \\
\Im{\rm m}\, (X\mathcal{V}^{M}F^{M+})  
& = & 
4 \mathcal{M}_{MN} \Re{\rm e} (X\mathcal{V}^{M}) \Im{\rm m}\,
(\mathcal{V}^{*N}\mathcal{V}_{M} F^{M+})\, ,
\\
& & \nonumber \\
\Im{\rm m}\, (\mathcal{V}^{M}\Phi^{x})  
& = & 
-2\mathcal{M}^{M}{}_{N}\Re{\rm e}\, (\mathcal{V}^{N}\Phi^{x})\, ,
\end{eqnarray}

\noindent
where

\begin{equation}
\mathcal{M}_{MN}   
\equiv 
\Re{\rm e}\, (\mathcal{V}^{*}_{M}\mathcal{V}_{N})
\end{equation}

\noindent
and using Eq.~(\ref{eq:bigidentity}) and the on-shell conditions
$\tilde{G}_{\Lambda}=\tilde{F}_{\Lambda}$ we can rewrite our previous result,
again to lowest order in fermions, in the form

\begin{equation}
  \begin{array}{rcl}
\left[\delta_{\eta},\delta_{\epsilon}\right] B_{0\, \mu\nu} 
 & = &   
2\partial_{[\mu}(-2\beta\xi_{\nu]}) 
-16\beta \mathcal{M}_{MN} \Re{\rm e}\, (X\mathcal{V}^{M})F^{N}{}_{\mu\nu}
\\
& & \\
& & 
+2\beta \alpha^{M} \mathcal{M}_{MN}  \vartheta^{x}
\Re{\rm e}(\mathcal{V}^{N} \Phi^{x}{}_{\mu\nu})\, ,
\\  
\end{array}
\end{equation}

\noindent
which involves quantities that have appeared in other commutators.

The last term in the r.h.s.~is a 2-form
St\"uckelberg shift. Its presence confirms that $\vartheta^{0}=0$ because
$F^{M}$ could never be gauge invariant containing $\vartheta^{0}B_{0}$. On the
other hand, this modification of the generic tensor hierarchy prediction is
possible because the generic case one does not take into account the possible
vanishing of components of the embedding tensor.

To make progress we need to identify the field strength $H_{0}$ of $B_{0}$,
which cannot have the trivial form predicted by the tensor hierarchy. The
r.h.s.~of the commutator suggests that $H_{0}$ must contain a term of the form
$\mathcal{M}_{MN}A^{M}\wedge dA^{N}$ and a coupling to the 3-forms
$C_{x}{}^{M}$, which, as we will see, are the only 3-forms available. The only
non-trivial possibility turns out to be

\begin{equation}
H_{0} =dB_{0} +\mathcal{M}_{MN}
\left[\, A^{M}\wedge dA^{N} +\alpha^{M}\vartheta^{x}C_{x}{}^{M}\, \right]\, ,
\end{equation}

\noindent
where $B_{0}$ must transform according to 

\begin{equation}
\label{eq:deltagaugeB0}
\delta B_{0\, \mu\nu}   
=
2\partial_{[\mu|}\Lambda_{0\, |\nu]}
+\mathcal{M}_{MN}
\left[\, 2A^{M}{}_{[\mu}\partial_{\nu]}\Lambda^{N} 
-\alpha^{M}\vartheta^{x}(\Lambda_{x}{}^{N}{}_{\mu\nu}
+\Lambda^{N}B_{x\, \mu\nu} +\Lambda_{x\, [\mu}A^{N}{}_{\nu]})\, \right]\, .
\end{equation}

Then, we see that ($\beta =-1/8$) the supersymmetry transformation 

\begin{equation}
\label{eq:susyB0}
\delta_{\epsilon}B_{0\, \mu\nu}
= 
-\tfrac{i}{8} 
\bar{\epsilon}^{I}\gamma_{[\mu}\psi_{\nu] I} +\mathrm{c.c.}
+2\mathcal{M}_{MN} A^{M}{}_{[\mu} 
\delta_{\epsilon} A^{N}{}_{\nu]}
\, ,
\end{equation}

\noindent
closes to all order in fermion fields if we impose the duality (on-shell)
condition

\begin{equation}
\tilde{H}_{0}=0\, ,  
\end{equation}

\noindent
where 

\begin{equation}
\tilde{H}_{0\, \mu\nu\rho}
=
H_{0\, \mu\nu\rho}
+\tfrac{3i}{8}\bar{\psi}^{I}{}_{[\mu}\gamma_{\nu| }\psi_{I\, |\rho]} \, .
\end{equation}

\noindent
This on-shell condition can also be rewritten as a duality between the bosonic
3-form field strength $H_{0}$ and the Noether current 1-form $j_{N\, 0}$
associated to the invariance under the global $U(1)_{\rm Kahler}$.

The 1-form parameter $\Lambda_{0}$ is given by

\begin{equation}
\Lambda_{0\, \mu} \equiv   
-\tfrac{1}{4}\xi_{\mu} -b_{0\, \mu}
+\mathcal{M}_{MN}\left[\, a^{M} +2 \Re{\rm
    e}(X\mathcal{V}^{M})\, \right]A^{N}{}_{\mu} \, .
\end{equation}

Again, the consistency of these results relies on the existence of the
appropriate 3-forms, which we explore next.


\subsection{3-forms}

The only supersymmetry transformations for 3-forms that lead to closure of the
local supersymmetry algebra are

\begin{equation}
\delta_{\epsilon}  C_{x}{}^{M}{}_{\mu\nu\rho}
= -\tfrac{3}{8}\mathcal{V}^{*M}\sigma^{x}{}_{I}{}^{J}\varepsilon^{IK}
\bar{\epsilon}_{K}\gamma_{[\mu\nu}\psi_{\rho]\, J} +\mathrm{c.c.}
-3\delta_{\epsilon}A^{M}{}_{[\mu|}B_{x| \nu\rho]}\, .
\end{equation}

\noindent
In particular, it is easy see that the supersymmetry algebra does not close
with

\begin{equation}
\delta_{\epsilon}  C_{0}{}^{M}{}_{\mu\nu\rho}
= \lambda \mathcal{V}^{*M}\varepsilon^{IJ}
\bar{\epsilon}_{I}\gamma_{[\mu\nu}\psi_{\rho]\, J} +\mathrm{c.c.}\, ,
\end{equation}

\noindent
for any values of $\lambda$, just as it did not close on the candidate $A^{x}$
and we conclude that $C_{0}{}^{M}$ cannot be introduced in this theory, which
agrees with the fact that a $\vartheta^{0}$ cannot be introduced, either.

The closure of the local supersymmetry algebra requires the use of the
previously found on-shell conditions $\tilde{F}^{M}=\tilde{G}^{M}$ and
$\textit{H}_{x}=0$ and a of the new condition

\begin{equation}
\tilde{G}_{x}{}^{M} =
-\tfrac{3}{4} \star \mathcal{M}^{MN}\alpha_{N}\vartheta^{x}\, ,
\end{equation}

\noindent
where the supercovariant 4-form field strength $\tilde{G}_{x}{}^{M}$ is given
by

\begin{equation}
\tilde{G}_{x}{}^{M} =
G_{x}{}^{M} -\tfrac{3}{4}
\left[\mathcal{V}^{M}\sigma^{x}{}_{I}{}^{K}\varepsilon_{JK} 
\bar{\psi}^{I}{}_{[\mu}\gamma_{\nu\rho}\psi^{J}{}_{\sigma]}
+\mathrm{c.c.}
\right]\, . 
\end{equation}

The on-shell condition is the supersymmetrization of the one proposed in
Ref.~\cite{Bergshoeff:2009ph}

\begin{equation}
G_{x}{}^{M} =\tfrac{1}{2} 
\star \frac{\partial V}{\partial \vartheta_{M}{}^{x}}\, ,  
\end{equation}

\noindent
for a manifestly symplectic-invariant (constant) potential $V$ given by 

\begin{equation}
V= -\tfrac{3}{4}
\mathcal{M}^{MN}\alpha_{M}\alpha_{N}\vartheta^{x}\vartheta^{x}\, ,
\end{equation}

\noindent
which generalizes the standard one.

Finally, the 3-form gauge parameter is given by the bilinear

\begin{equation}
\label{eq:L1x+L2x}
\Lambda_{x\, \mu\nu\rho} 
 =  
+\tfrac{3}{16}(\star \xi^{x})_{\mu\nu\rho}
+\tfrac{3}{8}\varepsilon_{xyz}B_{y\, [\mu\nu|}b_{z\, |\rho]}
+\tfrac{1}{2}a_{P}C_{x}{}^{P}{}_{\mu\nu\rho}
-d_{x\, \mu\nu\rho}\, ,
\hspace{1cm}
d_{x\, \mu\nu\rho}
\equiv   
\xi^{\sigma}D_{x\, \sigma\mu\nu\rho}\, .
\end{equation}


\subsection{4-forms}

There are three candidates to supersymmetry transformation rules of
4-forms\footnote{The transformation $\bar{\epsilon}^{I}
  \gamma_{[\mu\nu\rho}\psi_{\sigma]\, I}+\mathrm{c.c.}$ is the transformation
  of the the volume 4-form.}:

\begin{eqnarray}
\delta_{\epsilon}  D^{\prime}_{x\, \mu\nu\rho\sigma} 
& = & 
 \sigma^{x}{}_{I}{}^{J}\bar{\epsilon}^{I} 
\gamma_{[\mu\nu\rho}\psi_{\sigma]\, J}+\mathrm{c.c.}\, ,
\\
& & \nonumber \\  
\delta_{\epsilon}D^{\prime\prime}_{x\, \mu\nu\rho\sigma}
& = &  
C_{x\, M\, \mu\nu\rho|}\delta_{\epsilon} A^{M}{}_{|\sigma]}\, ,
\\
& & \nonumber \\  
\delta_{\epsilon}D^{\prime\prime\prime}_{x\, \mu\nu\rho\sigma}
& = &  
\varepsilon_{xyz} B_{y\, [\mu\nu|}\delta_{\epsilon}
B_{z|\rho\sigma]}\, .
\end{eqnarray}

Let us first consider the ungauged case, for simplicity. In this case, the
4-forms decouple from the rest of the hierarchy and the gauge transformations
may differ from those derived in the gauged case. The commutator of two
supersymmetries closes for all three candidates at the lowest order on
fermions, which would contradict the prediction made in
Ref.~\cite{Kleinschmidt:2008jj} in the framework of the KM approach that there
are only two triplets of 4-forms. Thus, we are lead to study the quartic terms
in fermions in the r.h.s.~of the commutators, as in
Ref.~\cite{Bergshoeff:2010mv}. These terms, which do not correspond to any of
the gauge parameters found for the lower-rank forms, do not vanish for any of
the three candidates and we can do two different things about it:

\begin{enumerate}
\item We can define as many 4-form gauge parameters as quartic terms we
  find. There are four different quartic terms and they appear in the
  commutator of the three candidates, but always in a fixed combination so
  there is, actually, only one independent 4-form gauge parameter. This gauge
  parameter may be used to gauge away one of the three candidates, which would
  leave us with the two independent triplets predicted by the KM approach.

\item We can construct linear combinations

\begin{equation}
aD^{\prime}_{x}+bD^{\prime\prime}_{x}+cD^{\prime\prime\prime}_{x}\, ,
\end{equation}

of the three (triplets) of candidate 4-forms and choose the coefficients so
that the quartic terms vanish. Since they always appear in the same combination
we get only one constraint for the coefficients $a,b,c$

\begin{equation}
6a+\frac{3}{8}b+\frac{1}{2}c=0\, ,  
\end{equation}

\noindent
which leaves us with two independent combinations on which the supersymmetry
algebra, again in perfect agreement with the KM approach prediction.

\end{enumerate}

Let us now consider the gauged case. The r.h.s.~of the commutator contains new
terms quadratic in fermions but the same terms quartic in fermions, so the
above discussion still applies. All the quadratic terms but two correspond to
gauge parameters already defined. The two terms that do not are the total
derivative of a 3-form and a 4-form shift. If the latter does not cancel, we
can gauge away one more triplet, so only one would remain. To cancel it, we
must have

\begin{equation}
c=\frac{3}{8}b\, ,  
\end{equation}

\noindent
which also leaves us with only one possible triplet of 4-forms, up to overall
normalization.  The overall normalization is fixed by the requirement that the
remaining combination coincides with the triplet of 4-forms predicted by the
tensor hierarchy $a=-3/16$.

Summarizing: in the gauged case new St\"uckelberg symmetries appear which
leave use with only one triplet which is the one predicted by the tensor
hierarchy and transforms under supersymmetry according to

\begin{equation}
\delta_{\epsilon}D_{x\, \mu\nu\rho\sigma}
= 
-\tfrac{3}{16} \sigma^{x}{}_{I}{}^{J}\bar{\epsilon}^{I} 
\gamma_{[\mu\nu\rho}\psi_{\sigma]\, J}+\mathrm{c.c.}  
+2 C_{x\, M\, [\mu\nu\rho|}\delta_{\epsilon} A^{M}{}_{|\sigma]}
+\tfrac{3}{4} \varepsilon_{xyz} B_{y\, [\mu\nu|}\delta_{\epsilon}
B_{z|\rho\sigma]}\, .
\end{equation}

The supersymmetry algebra closes with the gauge transformations
Eq.~(\ref{eq:deltahDx1+Dx2}) plus the St\"uckelberg shift

\begin{equation}
\tilde{\Lambda}_{\mu\nu\rho\sigma}
=
-\tfrac{9}{4}\alpha^{M}\mathcal{M}_{MN}\Re{\rm e}\, (X\mathcal{V}^{N}) 
\varepsilon_{\mu\nu\rho\sigma}   
+\alpha_{M}b_{y\, [\mu}[C_{y}{}^{M}+3A^{M}B_{y}]_{|\nu\rho\sigma]}\, ,
\end{equation}

\noindent
which could be used to eliminate one component of the triplet, at the expense
of breaking explicit $SU(2)$-invariance.

We conclude that the tensor hierarchy of pure, gauged, $N=2,d=4$ supergravity
contains two 1-forms which are $SU(2)$ singlets, $A^{M}$, four 2-forms (a
triplet and a singlet) $B_{x}$ and $B_{0}$, six 3-forms (two $SU(2)$ triplets)
$C_{x}{}^{M}$ and one triplet of 4-forms $D_{x}$ (two in the ungauged
case). It seems that the predictions of the KM approach have to be modified
after gauging.


\section{Pure $N=2,d=5$ supergravity}
\label{sec-5dtheory}

In this section we are going to study the case of pure minimal supergravity in
5 dimensions \cite{Cremmer:1980gs}. The possible $(p+1)$ forms that can be
coupled to the ungauged theory consistently with supersymmetry have been
studied in Refs.~\cite{Gomis:2007gb} and \cite{Kleinschmidt:2008jj}. Here we
are going to revise those results taking into account the predictions of the
generic 5-dimensional tensor hierarchy constructed in \cite{Hartong:2009vc}
using as global symmetry group the R-symmetry group $SU(2)$ (gauge invariance
is clearly a pre-condition for supersymmetry invariance).

The supergravity multiplet of the $N=2,d=5$ theory consists of the graviton
$e^{a}{}_{\mu}$, a symplectic-Majorana gravitino $\psi_{I\, \mu}\, ,\,\,\,
(I=1,2)$ and one graviphoton $A_{\mu}$. 

The global symmetry group of this theory (or its equations of motion) reduces
to the $SU(2)$ R-symmetry group and so the embedding tensor is
$\vartheta^{x}\, ,\,\,\,\, x=1,2,3$. In the standard formulations of the
theory it appears as a Fayet-Iliopoulos term that selects the $U(1)$ subgroup
of $SU(2)$ which is going to be gauged by the graviphoton. The formulation of
the gauged theory in terms of the fundamental fields is, therefore, well
known\footnote{See \cite{Bellorin:2007yp}, whose conventions we follow, and
  references therein.}: the bosonic action of the fundamental fields
is given by\footnote{The constant $c$ stands for the unique components of the
  totally-symmetric tensor $C_{IJK}$.}

\begin{equation}
\label{eq:pureungaugedN2D5bosonicaction}
  S  =  {\displaystyle\int}   \left[\star R 
+\tfrac{1}{2}c^{2/3}F\wedge \star F +\tfrac{1}{3\sqrt{3}}c F\wedge F
    \wedge A -\star V \right]\, ,
\end{equation}

\noindent
where 

\begin{equation}
F=dA\, ,
\hspace{1cm}
V=-4c^{-2/3}\vartheta^{x}\vartheta^{x}\, ,  
\end{equation}

\noindent
and where the supersymmetry transformations of the fundamental fields to all
order in fermions are

\begin{eqnarray}
\delta_{\epsilon} e^{a}{}_{\mu} 
& = & 
{\textstyle\frac{i}{2}} \bar{\epsilon}_{i}\gamma^a\psi^{i}_{\mu}\, ,
\\
& & \nonumber \\ 
\delta_{\epsilon} A_{\mu} 
& = & 
-{\textstyle\frac{i\sqrt{3}}{2}}c^{-1/3}\bar{\epsilon}_{i}\psi^{i}_{\mu}\, ,
\\
& & \nonumber \\ 
\delta_{\epsilon}\psi^{i}_{\mu}\, , 
& = & 
\tilde{\mathfrak{D}}_{\mu}\epsilon^{i}
-{\textstyle\frac{1}{8\sqrt{3}}}c^{1/3}\tilde{F}^{\alpha\beta}
\left(\gamma_{\mu\alpha\beta}-4g_{\mu\alpha}\gamma_\beta\right)
\epsilon^{i}
+\tfrac{i}{2\sqrt{3}}c^{-1/3}\vartheta^{x} \sigma^{x\, i}{}_{j} 
\gamma_{\mu}\epsilon^{j}\, .
\end{eqnarray}

\noindent
The covariant derivative is given by

\begin{equation}
  \tilde{\mathfrak{D}}_{\mu}\epsilon^{i} \equiv \tilde{\nabla}_{\mu}\epsilon^{i} 
  +\tfrac{i}{2}A_{\mu}\vartheta^{x} \sigma^{x\, i}{}_{j} \epsilon^{j}\, ,
\end{equation}

\noindent
where $\tilde{\nabla}_{\mu}$ is the Lorentz-covariant derivative with the
torsionful connection $\tilde{\omega}_{\mu}{}^{ab}$ defined in
Eq.~(\ref{eq:torsionfuld4}) where the torsion is now given by

\begin{equation}
T_{\mu\nu}{}^{a} \equiv -\tfrac{i}{2} \bar{\psi}_{i\,
  [\mu}\gamma^{a}\psi^{i}{}_{\nu]}\, .
\end{equation}

$\tilde{F}_{\mu\nu}$ is the supercovariant 2-form field strength:

\begin{equation}
\tilde{F}_{\mu\nu} \equiv F_{\mu\nu} 
+\tfrac{i\sqrt{3}}{2} c^{-1/3}
\bar{\psi}_{i\, \mu}\psi^{i}{}_{\nu}\, .    
\end{equation}

Following \cite{Hartong:2009vc} one finds that the tensor hierarchy contains,
in addition to the graviphoton $A$, its dual 2-form $B$, a triplet of 3-forms
$C_{x}$ a triplet of 4-forms $D_{x}$, and, possibly, one 5-form $E$ which may
be consistently eliminated from the hierarchy if the 4-form gauge parameter
$\Lambda^{(4)}$ vanishes:

\begin{eqnarray}
\delta_{h}A 
& = & 
-d\Lambda^{(0)}\, ,
\\
& & \nonumber \\
\delta_{h}B 
& = & 
d\Lambda^{(1)} 
+\tfrac{1}{\sqrt{3}}c\Lambda^{(0)}F -\vartheta^{x}\Lambda^{(2)}{}_{x}\, ,
\\
& & \nonumber \\
\delta_{h}C_{x} 
& = & 
\mathfrak{D}\Lambda^{(2)}{}_{x}
+\varepsilon_{xyz}\vartheta^{y}(\Lambda^{(0)}C_{z} -\Lambda^{(3)}{}_{z})\, ,
\\
& & \nonumber \\
\delta_{h}D_{x} 
& = & 
\mathfrak{D}\Lambda^{(3)}{}_{x}
-F\wedge \Lambda^{(2)}{}_{x} 
+\varepsilon_{xyz}\vartheta^{y}\Lambda^{(0)}D_{z}
-\vartheta^{x} \Lambda^{(4)}\, ,
\\
& & \nonumber \\
\delta_{h}E
& = & 
d\Lambda^{(4)}\, .
\end{eqnarray}

\noindent
The corresponding gauge-covariant field strengths are

\begin{eqnarray}
F 
& = & 
dA\, ,
\\
& & \nonumber \\
H 
& = & 
dB +\tfrac{1}{\sqrt{3}}cA\wedge F +\vartheta^{x}C_{x}\, ,
\\
& & \nonumber \\
G_{x} 
& = & 
\mathfrak{D}C_{x} +\varepsilon_{xyz}\vartheta^{y}D_{z}\, ,
\\
& & \nonumber \\
K_{x} 
& = & 
\mathfrak{D}D_{x} +F\wedge C_{x}+\vartheta^{x}E\, ,
\\
& & \nonumber \\
L 
& = & 
dE\, .
\end{eqnarray}

In Ref.~\cite{Gomis:2007gb} it was shown that there is no independent 5-form
$E$ that can be introduced in the supersymmetric theory (consistently with
their find that $\Lambda^{(4)}=0 $) while in Ref.~\cite{Kleinschmidt:2008jj}
it was shown that the ungauged supersymmetric theory may admit an independent
triplet of 5-forms which should be decoupled from the rest of the tensor
hierarchy even in the gauged case, according to the above results. In order to
clarify these points we are going to construct the supersymmetric tensor
hierarchy of the gauged theory. Closing the supersymmetry algebra on the
different fields of the tensor hierarchy we will find the values of the gauge
parameters and we will be able to determine the necessity or impossibility of
adding 5-forms to it.

The supersymmetry algebra closes on the graviphoton giving

\begin{equation}
\left[ \delta_{\eta},\delta_{\epsilon}\right] A_{\mu} = 
\delta_{h} A_{\mu} + \delta_{\xi} A_{\mu}+ \delta_{\kappa} A_{\mu}\, ,
\end{equation}

\noindent
where $\delta_{h} A_{\mu}$ is the above gauge transformation for $A_{\mu}$
with gauge parameter

\begin{equation}
\Lambda^{(0)} 
=  
\lambda^{(0)} +\xi^{\mu}A_{\mu}\, ,
\hspace{1cm}
\lambda^{(0)}
 \equiv  
\tfrac{\sqrt{3}i}{2}c^{-1/3}\bar{\epsilon}_{i}\eta^{i}\, ,
\end{equation}

\noindent
and $\delta_{\xi}$ is a general coordinate transformation ($\pounds_{\xi}$)
with parameter

\begin{equation}
\xi_{\mu} 
=
\tfrac{i}{2}\bar{\epsilon}_{i}\gamma_{\mu}\eta^{i}\, ,
\end{equation}

\noindent
and $\delta_{\kappa} A_{\mu}$ is a supersymmetry transformation with parameter

\begin{equation}
\kappa^{i} = -\xi^{\mu}\psi^{i}{}_{\mu}\, .  
\end{equation}

For the 2-form we find that the supersymmetry algebra closes for the
supersymmetry transformation

\begin{equation}
\delta_{\epsilon}B_{\mu\nu}
= -i\sqrt{3}c^{1/3}\bar{\epsilon}_{i}\gamma_{[\mu}\psi^{i}_{\nu]} 
+ \tfrac{2}{\sqrt{3}}c A_{[\mu}\delta_{\epsilon}A_{\nu]}\, ,
\end{equation}

\noindent
with the parameters of the tensor hierarchy gauge transformations being given
by 

\begin{eqnarray}
\Lambda^{(1)}{}_{\mu}
& = & 
\sqrt{3}c^{1/3} \xi_{\mu} -\xi^{\rho}B_{\rho\mu} +\tfrac{1}{\sqrt{3}}
c\lambda^{(0)}A_{\mu}\, ,
\\
& & \nonumber \\
\Lambda^{(2)}{}_{x\, \mu\nu} 
& = & 
\lambda^{(2)}{}_{x\, \mu\nu} -\xi^{\rho}C_{x\, \rho\mu\nu}\, ,
\hspace{1cm}
\lambda^{(2)}{}_{x\, \mu\nu}
\equiv 
\tfrac{1}{2} \sigma^{x\, k}{}_{j}\varepsilon_{ik}\bar{\epsilon}^{i}\gamma_{\mu\nu}\eta^{j}\, ,
\end{eqnarray}

\noindent
if the supercovariant field strength 3-form is given by  

\begin{equation}
\tilde{H}_{\mu\nu\rho} \equiv H_{\mu\nu\rho} +\tfrac{i3\sqrt{3}}{2}c^{1/3}
\bar{\psi}_{i\, [\mu}\gamma_{\nu}\psi^{i}{}_{\nu]}\, ,  
\end{equation}

\noindent
($H$ being as predicted by the tensor hierarchy) the duality relation

\begin{equation}
\tilde{H}=c^{2/3}\star \tilde{F}\, ,  
\end{equation}

\noindent
is satisfied.

The supersymmetry algebra also closes in the 3-form with supersymmetry
transformations

\begin{equation}
\delta_{\epsilon} C_{x\, \mu\nu\rho} 
 =  
-\tfrac{3i}{2}
\sigma^{x}{}_{i}{}^{k} \varepsilon_{jk}
\bar{\epsilon}^{i}\gamma_{[\mu\nu}\psi^{j}_{\rho]}\, ,
\end{equation}

\noindent
with  the parameters of the gauge transformations predicted by the tensor
hierarchy being given by

\begin{equation}
\Lambda^{(3)}{}_{x\, \mu\nu\rho}
= 
\lambda^{(3)}{}_{x\, \mu\nu\rho} -\xi^{\sigma}D_{x\, \sigma \mu\nu\rho}
+\lambda^{(0)}C_{x\, \mu\nu\rho}\, ,
\hspace{1cm}
\lambda^{(3)}{}_{x\, \mu\nu\rho}
\equiv 
\tfrac{\sqrt{3}}{2}c^{-1/3} 
\sigma^{x\, k}{}_{j}\varepsilon_{ik}\bar{\epsilon}^{i}\gamma_{\mu\nu\rho}\eta^{j}\, ,
\end{equation}

\noindent
if the on-shell condition

\begin{equation}
\tilde{G}_{x\, \mu\nu\rho\sigma}
\equiv 
G_{x\, \mu\nu\rho\sigma} 
+3i\sigma^{x}{}_{i}{}^{k} \varepsilon_{jk}
\bar{\psi}^{i}{}_{[\mu}\gamma_{\nu\rho}\psi^{j}_{\sigma]} = 0\, ,  
\end{equation}

\noindent
is satisfied (with the same interpretation as the on-shell condition
$\tilde{H}_{\mathbf{x}}=0$ in the 4-dimensional case), and on the 4-forms
$D_{x}$ with

\begin{eqnarray}
\delta_{\epsilon} D_{x\, \mu\nu\rho\sigma} 
=  
2\sqrt{3}c^{-1/3}\sigma^{x}{}_{i}{}^{k} \epsilon_{jk} 
\bar{\epsilon}^{i} \gamma_{[\mu\nu\rho}\psi^{j}_{\sigma]} 
+4 C_{x\, [\mu\nu\rho}\delta_{\epsilon} A_{\sigma]}\, ,
\end{eqnarray}

\noindent
with

\begin{equation}
\Lambda^{(4)}{}_{\mu\nu\rho\sigma}  
= -\xi^{\delta}E_{\delta\mu\nu\rho\sigma}\, ,
\end{equation}

\noindent
if the (on-shell) duality relation

\begin{equation}
\star\tilde{K}_{x}=
-4c^{-2/3}\vartheta^{x}=\tfrac{1}{2}\frac{\partial V}{\partial \vartheta^{x}}\, ,  
\end{equation}

\noindent
is also satisfied, with

\begin{equation}
\tilde{K}_{x\, \mu_{1}\cdots \mu_{5}} \equiv K_{x\, \mu_{1}\cdots \mu_{5}}
-5\sqrt{3}c^{-1/3}\sigma^{x}{}_{i}{}^{k} \epsilon_{jk} 
\bar{\psi}^{i}{}_{[\mu} \gamma_{\nu\rho\sigma}\psi^{j}_{\lambda]}\, .   
\end{equation}

The 4-form shift $\Lambda^{(4)}{}_{x}$ appears only because we have included a
5-form $E$ in $K_{x}$. However, as discussed in Ref.~\cite{Gomis:2007gb}, a
non-trivial 5-form singlet cannot be introduced in the theory. Therefore, The
supersymmetric tensor hierarchy seems to stop at the 4-form level and there is
no 5-form singlet $E$. However, in Ref.~\cite{Kleinschmidt:2008jj} it was
shown that, in the ungauged case, as predicted by the KM approach, the
supersymmetry algebra closes for a triplet of 5-forms with supersymmetry
transformations of the form

\begin{equation}
\delta_{\epsilon}E_{x\, \mu_{1}\cdot \mu_{5}} 
= B_{[\mu_{1}\mu_{2}}\delta_{\epsilon}C_{|x|\, \mu_{3}\mu_{4}\mu_{5}]}
+\tfrac{1}{2\sqrt{3}}c 
A_{[\mu_{1}}\delta_{\epsilon}D_{|x|\, \mu_{2}\mu_{3}\mu_{4}\mu_{5}]}\, ,
\end{equation}

\noindent
up to an overall normalization constant which cannot be fixed because this
triplet does not couple to the rest of the fields of the tensor hierarchy.  We
have checked the closure of the supersymmetry algebra to all orders in
fermions for this triplet in the ungauged case but in the gauged case the
closure takes place only up to 5-form St\"uckelberg shifts proportional to
$\vartheta^{x}\Lambda^{(5)}$ and $\varepsilon_{xyz}\vartheta^{y}
\Lambda{}^{(5)}{}_{z}$ which, on the one hand, cannot be compensated by those
of any other 5-forms and, on the other hand, can be used to gauge away the
full $E_{x}$, recovering the spectrum of higher-rank forms predicted by the
tensor hierarchy, as in the 4-dimensional case.


\section{Pure $N=(2,0),d=6$ supergravity}
\label{sec-6dtheory}

Let us now consider pure $N=(2,0), d=6$ supergravity \cite{Romans:1986er}. We
shall be extremely brief. The supergravity multiplet consists of the graviton
$e^{a}{}_{\mu}$, a positive-chirality symplectic-Majorana-Weyl gravitino
$\psi_{\mu I}$ and a 2-form $B_{\mu\nu}$ with field strength $H=dB$ and with
supercovariant field strength that is constrained to be self-dual
(i.e.~$\tilde{H}^{-}=0$).

The bosonic equations of motion, which cannot be derived from a
Lorentz-covariant Lagrangian are

\begin{eqnarray}
R_{\mu\nu} - \tfrac{1}{2}g_{\mu\nu}R -H^{\rho\sigma}_{\mu}H_{\nu\rho\sigma}
& = & 0\, ,\\
& & \nonumber \\
H^{-} & = &
0\, .
\end{eqnarray}

The supersymmetry transformations for the supergravity fields are:

\begin{eqnarray}
\delta_{\epsilon}e^{a}{}_{\mu}
& =& 
\bar{\psi}^{I}{}_{\mu}\gamma^{a} \epsilon_{I}\, ,
\\
& & \nonumber \\ 
\delta_{\epsilon}\psi_{I\, \mu}
& = & 
\tilde{\nabla}_{\mu}\epsilon_{I} 
+\tfrac{1}{4}\tilde{H}_{\mu\nu\rho}\gamma^{\nu\rho}\epsilon_{I}\, ,
\\
& & \nonumber \\ 
\delta_{\epsilon}B_{\mu\nu} 
& = & 
\bar{\epsilon}^{I}\gamma_{[\mu}\psi_{\nu ] I}\, . 
\end{eqnarray}

The global symmetry group of this theory reduces to the $SU(2)$ R-symmetry
group but there are no 1-forms available to gauge them and the embedding
tensor vanishes. The 6-dimensional tensor hierarchy allows for deformations
which are not gaugings and depend on deformation parameters which do not depend
on the embedding tensor\footnote{Supersymmetry constraints may modify this
  last statement.} \cite{Hartong:2009az}. However, the absence of 1-forms
implies the absence of 3-forms in the tensor hierarchy and the vanishing of
any other deformation parameters and non-trivial constraints between
them. This implies, according to the general arguments of
Ref.~\cite{Hartong:2009az} the absence of 5- and 6-forms in the tensor
hierarchy. The only higher-rank forms allowed would be a triplet of 4-forms
$D_{x}$ related to the Noether currents of the $SU(2)$ $R$-symmetry group
which are bilinear in gravitini. The tensor hierarchy would, then, consist of
$\{B_{\mu\nu},D_{x\, \mu\nu\rho\sigma}\}$ with trivial gauge transformations and
field strengths

\begin{equation}
H\equiv dB\, ,
\hspace{1cm}
K_{x}\equiv d D_{x}\, .   
\end{equation}

It is not difficult to see that supersymmetry confirms these results: it is
not possible to construct consistent supersymmetry transformations for any
other kind of higher-rank fields (except for the one corresponding to the
volume 6-form).

In the ungauged, though, it is possible to construct consistent supersymmetry
transformation for a triplet of 6-forms $F_{x}$:

\begin{equation}
\delta_{\epsilon}F_{x} =  B_{[\mu_{1}\mu_{2}|}\delta_{\epsilon}D_{x\,
  |\mu_{3}\cdots \mu_{6}]}\, , 
\end{equation}

\noindent
in agreement with the predictions of the KM approach
\cite{Kleinschmidt:2008jj}.


\section{Extended objects and their effective actions}
\label{sec-puren2objectsandactions}

Our goal in this section is to discuss the consistency of our previous
results: we will study the relations between the fields of the $d=4,5,6$
tensor hierarchies through dimensional reduction and the possible existence of
supersymmetric branes charged with respect to them, constructing the bosonic
parts of some of their worldvolume actions.

\begin{figure}[!ht]
\begin{center}
\leavevmode
  \includegraphics[scale=1]{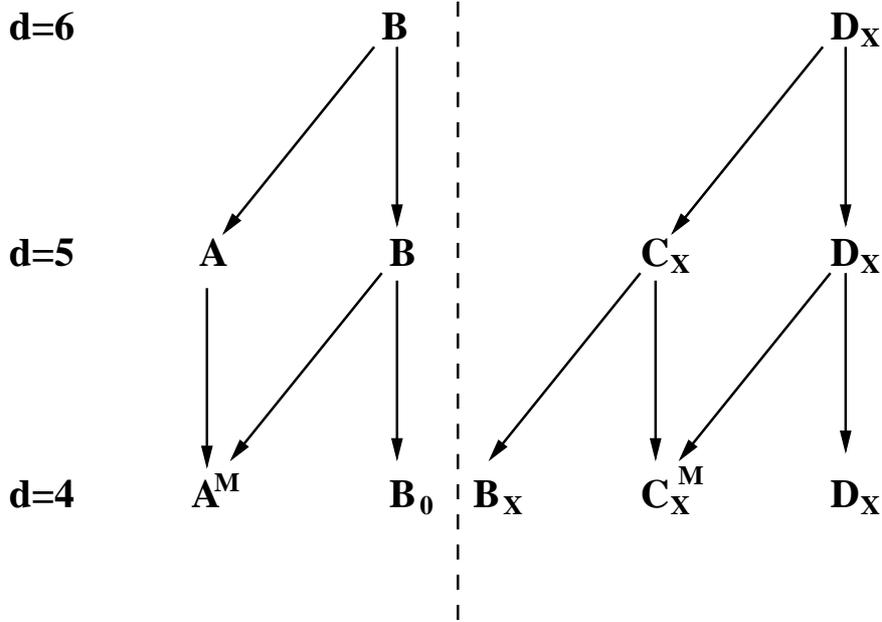}
  \caption{Relations between the fields of the tensor hierarchies of the
    $d=4,5,6$ pure supergravities with 8 supercharges. The fields that descend
    from the 6-dimensional 4-forms $D_{x}$, to the right of the dashed line,
    are associated to worldvolume theories with 2 bosonic and 2 fermionic (8
    divided by 2 ($\kappa$-symmetry) and by 2 (e.o.m.)) degrees of
    freedom. The fields that descend from the 6-dimensional 2-form $B$ with
    selfdual 3-form field strength are associated to worldvolume theories with
    1 bosonic (right- or left movers for $p=1$) and 1 fermionic degrees of
    freedom.}
\label{fig:d456forms}
\end{center}
\end{figure}

The 3 pure supergravity theories with 8 supercharges that we have considered
are related by dimensional reduction and truncation of the extra vector
multiplet that appears in each reduction\footnote{See,
  e.g.~Ref.~\cite{LozanoTellechea:2002pn}.}. Then, their higher-rank
potentials should also be related by dimensional reduction if we ignore the
additional 1-forms of the vector supermultiplets, that we truncate. That this
is the case can immediately be seen in Figure~\ref{fig:d456forms}. Observe
that, as usual, electric-magnetic pairs in 4 dimensions are associated to
rotations in 6 dimensions. Observe also that one additional form in any
dimension would imply the existence of related forms in the other two which
leads us to conclude that indeed we have obtained all the possible higher-rank
forms.

If there are supersymmetric extended objects charged with respect to these
forms, then they must also be related by simple and double dimensional
reductions. On the other hand, all the dynamical $p$-branes of these theories
must be charged with respect to the $(p+1)$-form potentials that we have
found, since these are the only potentials of the theory transforming into the
gravitini under supersymmetry\footnote{We expect this fact to remain true even
  after coupling to matter.}. It should, then, be possible to construct a
$\kappa$-symmetric action for each of them with a Wess-Zumino term lead by
the pullback corresponding $(p+1)$-form potential.

A necessary condition for the $\kappa$-symmetric actions to exist is
Bose-Fermi matching of worldvolume degrees of freedom
\cite{Achucarro:1987nc,Duff:1992hu}. The worldvolume theory of the
6-dimensional 3-brane (whose worldvolume has the simplest standard form) does
not need any additional worldvolume bosonic fields: the 2 bosonic degrees of
freedom associated to the transverse coordinates exactly match the 2 fermionic
degrees of freedom that result from dividing the 8 of the minimal spinor by
$2\times 2$ ($\kappa$-symmetry and fermionic equations of motion). Reducing
to $d=5$ we get 2-branes and 3-branes. The theory of the latter contains an
additional worldvolume scalar (or, equivalently, a worldvolume
2-form). 

Simple dimensional reduction of the 5-dimensional 3-brane yields the
worldvolume theory of another 3-brane (now spacetime-filling) with two
non-geometrical scalars (or 2-forms). Double dimensional reduction gives a
domain wall with one additional scalar which can be dualized into a vector
which we expect to be of Born-Infeld type. The simple dimensional reduction of
the 5-dimensional 2-brane gives essentially the same result (up to
electric-magnetic rotation) while the double gives a string with no additional
degrees of freedom coupling to $B_{x}$. All the theories obtained from the
6-dimensional 3-brane have the same number of bosonic and fermionic degrees of
freedom.

Things are different if we start from the 6-dimensional (self-dual)
string. Bose-Fermi matching can only be achieved between the left- or
right-moving transverse scalars (which are 4) and the 2 fermionic degrees of
freedom. This characteristic is inherited by the 5- and 4-dimensional strings
which one obtains by simple dimensional reduction and which must contain
additional, non-geometric, worldvolume scalars.

As we have seen, in each of the ungauged $d=4,5,6$ cases there is one
additional triplet of top forms. These triplets are obviously related by
dimensional reduction. None of them seems to couple to supersymmetric
spacetime-filling branes, at least within the framework of a conventional
worldvolume action.

\section{Conclusions}
\label{sec-conclusions}

In this paper we have taken a step towards the democratic formulation of
$d=4,5,6$ supergravity theories with 8 supercharges (often called $N=2$
theories), finding all the potentials that transform into the gravitini under
supersymmetry. In the 4-dimensional case, our results complement those
obtained by de Vroome and de Wit in Ref.~\cite{de Vroome:2007zd} and in the
5-dimensional case those of Kleinschmidt and Roest in
Ref.~\cite{Kleinschmidt:2008jj}.

We have seen that the predictions of the bosonic tensor hierarchies are
essentially satisfied in the supersymmetric case. In the 4-dimensional case we
find some differences due to the impossibility of gauging the $U(1)$ factor of
the R-symmetry group, which leads to an additional constraint of the embedding
tensor. However, we have also found that in the ungauged case there are more
fields (only top forms) than predicted by this approach. This is, on the other
hand, in complete agreement with the predictions of the KM approach. In the 4-
and 5-dimensional cases we have shown that, after gauging, the new top forms
can either be completely gauged away or must be combined with other fields due
to the appearance of new St\"uckelberg shifts that depend on the embedding
tensor. To determine completely the number of independent top forms it has
been crucial to study the closure of the supersymmetry algebra to all orders
in fermion fields, as in Ref.~\cite{Bergshoeff:2010mv}.

The determination and study of the top form fields of these theories was
precisely on of our goals. We have determined them but their physical meaning
is, though, not completely clear: on general grounds the top forms should be
associated to the constraints imposed on the embedding tensor. However, in the
4-dimensional case we have solved all those constraints and a
Lagrange-multiplier $D_{x}$ is not really needed. Furthermore, we find one
extra triplet in each dimension that does not fit into the tensor hierarchy.

On the other hand, the 4-dimensional top forms in the tensor hierarchy seem to
be associated to the possible supersymmetric truncations from $N=2$ to $N=1$
in $d=4$.  Supersymmetric truncations are not possible in $d=5,6$, in
agreement with the absence of top forms in the tensor hierarchy in those
dimensions. The additional triplets of top forms do not seem to be related to
truncations and their existence, albeit predicted by the KM approach, remains
mysterious.

It would be interesting to construct explicitly the worldvolume actions that
contain the potentials we have found and check their relations via dimensional
reduction. This would shed more light on the relations between the
corresponding supersymmetric objects in different dimensions and also on the
possible intersections between supersymmetric objects of the same theory.  It
would also be very interesting to construct the complete democratic
formulations of the $N=2,d=4,5,6$ supergravities with matter couplings with
the help of the tensor hierarchies. This is, at any rate a necessary step to
find the most general $N=2$ supergravity theories. However, it would also give
us a deeper understanding of these theories and very useful tools to work with
them.


\section*{Acknowledgments}

We have benefited from conversations with, comments by and questions from
D.~Roest, to whom we would like to express our gratitude.  This work has been
supported in part by the Spanish Ministry of Science and Education grants
FPA2006-00783 and FPA2009-07692, the Comunidad de Madrid grant HEPHACOS
S2009ESP-1473 and the Spanish Consolider-Ingenio 2010 program CPAN
CSD2007-00042.  TO wishes to express his gratitude to the CERN Theory Division
for its hospitality during some stages of this work and M.M.~Fern\'andez for
her permanent support.

\appendix

\section{Some formulae}
\label{app-conventions}

Using Eq.~(\ref{eq:LL}) and the normalization of the canonical symplectic
section Eq.~(\ref{eq:conicalnorm}) (taking into account that in this case
$\mathcal{L}^{\Lambda}$ and $\mathcal{M}_{\Lambda}$ stand for a single
number), we find that

\begin{eqnarray}
\label{eq:ImMbarupNup}
\Im{\rm m}(\mathcal{V}^{*M}\mathcal{V}^{N}) 
\!\! & = & \!\!
-\tfrac{1}{2}\Omega^{MN}\, ,  
\\
& & \nonumber \\
\label{eq:ReMbarupNup}
\Re{\rm e}(\mathcal{V}^{*M}\mathcal{V}^{N}) 
\!\! & = & \!\!
-\tfrac{1}{2}
\left(
  \begin{array}{cc}
\Im{\rm m}\,
\mathcal{N}^{\Lambda\Sigma}
&    
\Im{\rm m}\,
\mathcal{N}^{\Lambda\Omega} \Re{\rm e}\mathcal{N}_{\Omega\Sigma}
\\
& \\
\Re{\rm e}\mathcal{N}_{\Lambda\Omega}
\Im{\rm m}\, \mathcal{N}^{\Omega\Sigma}
&
\Im{\rm m}\,
\mathcal{N}_{\Lambda\Sigma} +\Re{\rm e}\mathcal{N}_{\Lambda\Omega}
\Im{\rm m}\,
\mathcal{N}^{\Omega\Gamma}\Re{\rm e}\mathcal{N}_{\Gamma\Sigma}
\\
  \end{array}
\right)\, ,
\end{eqnarray}

\noindent
and 

\begin{eqnarray}
\label{eq:ImMbarupNdown}
\Im{\rm m}(\mathcal{V}^{*M}\mathcal{V}_{N}) 
\!\! & = & \!\!
-\tfrac{1}{2}\delta^{M}{}_{N}\, ,  
\\
& & \nonumber \\
\label{eq:ReMbarupNdown}
\Re{\rm e}(\mathcal{V}^{*M}\mathcal{V}_{N}) 
\!\! & = & \!\!
-\tfrac{1}{2}
\left(
  \begin{array}{cc}
\Im{\rm m}\,
\mathcal{N}^{\Lambda\Omega} \Re{\rm e}\mathcal{N}_{\Omega\Sigma}
&    
-\Im{\rm m}\,
\mathcal{N}^{\Lambda\Sigma}
\\
& \\
\Im{\rm m}\,
\mathcal{N}_{\Lambda\Sigma} +\Re{\rm e}\mathcal{N}_{\Lambda\Omega}
\Im{\rm m}\,
\mathcal{N}^{\Omega\Gamma}\Re{\rm e}\mathcal{N}_{\Gamma\Sigma}
&
-\Re{\rm e}\mathcal{N}_{\Lambda\Omega}
\Im{\rm m}\, \mathcal{N}^{\Omega\Sigma}
\\
  \end{array}
\right)\, ,
\end{eqnarray}

\noindent
so, in particular,

\begin{equation}
\label{eq:RR}
\Re{\rm e}(\mathcal{V}^{*\, M}\mathcal{V}_{N}) 
\Re{\rm e}(\mathcal{V}^{*\, N}\mathcal{V}_{P})
=
-\tfrac{1}{4}\delta^{M}{}_{P}\, , 
\end{equation}

\noindent
and

\begin{equation}
\label{eq:bigidentity}
\Im{\rm m}(\mathcal{V}^{*M}\mathcal{V}_{N}F^{N\, +})=
-\tfrac{1}{2}F^{M}
-\tfrac{1}{4} 
\left(
  \begin{array}{c}
\Im{\rm m}\,
\mathcal{N}^{\Lambda\Omega} \star (G_{\Omega}-F_{\Omega})
\\
\\
(G_{\Lambda}-F_{\Lambda}) 
+
\Re{\rm e}\mathcal{N}_{\Lambda\Sigma}
\Im{\rm m}\, \mathcal{N}^{\Sigma\Omega}(G_{\Omega}-F_{\Omega})
\end{array}
\right)\, .
\end{equation}


\end{document}